# Electrochemical Properties of Cu(II/I)-Based Redox Mediators for Dye-Sensitized Solar Cells

Ladislav Kavan[1,2*], Yasemin Saygili[3], Marina Freitag[3,4], Shaik M. Zakeeruddin[2], Anders Hagfeldt[3] and Michael Grätzel[2]

[1]*J. Heyrovský Institute of Physical Chemistry, v.v.i., Academy of Sciences of the Czech Republic, Dolejškova 3, CZ-18223 Prague 8, Czech Republic*
[2]*Laboratory of Photonics and Interfaces, Institute of Chemical Sciences and Engineering, Swiss Federal Institute of Technology, CH-1015 Lausanne, Switzerland*
[3]*Laboratory of Photomolecular Science, Institute of Chemical Sciences and Engineering, Swiss Federal Institute of Technology, CH-1015, Lausanne, Switzerland*
[4]*Department of Chemistry, Ångström Laboratory, Uppsala University, 751 20 Uppsala, Sweden*

*e-mail: kavan@jh-inst.cas.cz

Three Cu(II/I)-phenanthroline and Cu(II/I)-bipyridine redox mediators are studied on various electrodes and in variety of electrolyte solutions using cyclic voltammetry and impedance spectroscopy on symmetrical dummy cells. Graphene-based catalysts provide comparably high activity to PEDOT, and both catalysts outperform the activity of platinum. The charge-transfer kinetics and the diffusion rate significantly slowdown in the presence 4-*tert*-butylpyridine. This effect is specific only for Cu-mediators (is missing for Co-mediators), and is ascribed to a sensitivity of the coordination sphere of the Cu(II)-species to structural and substitutional changes. The 'Zombie Cells' made from symmetrical PEDOT/PEDOT devices exhibit enhanced charge-transfer rate and enhanced diffusion resistance. Electrochemically clean Cu(II)-bipyridine species are prepared, for the first time, by electrochemical oxidation of the parent Cu(I) complexes. Our preparative electrolysis brings numerous advantages over the standard chemical syntheses of the Cu(II)-bipyridine complexes. The superior performance of electrochemically-grown clean Cu(II)-bipyridine complex is demonstrated on practical dye-sensitized solar cells.

# 1. Introduction

The dye sensitized solar cell (DSC) also called the Graetzel cell [1,2] offers high efficiency, low-cost, and unprecedented versatility in the optimization of its components, i.e. electrode materials, dyes and hole transporting media including electrolyte solutions. One of the key tasks in boosting the performance of the most common (liquid-junction) DSC is the enhancement of the cell's photovoltage through the selection of electrolyte redox mediators. The reason is that the open-circuit voltage ($V_{OC}$) of a DSC stems from the difference of the quasi-Fermi level position in the semiconductor photoanode and the energy level corresponding to the redox potential of the mediator [1-3]. Consequently, the enhancement of $V_{OC}$ can be realized by two general strategies: (i) downshifting of the Fermi level (in the scale of electrochemical potential) and (ii) upshifting of the mediators' redox potential.

The first strategy is mostly based on electrolyte additives such as 4-*tert*-butylpyridine (TBP) which causes not only the desired Fermi level shift, but also increases the resistance to electron recombination [4-6]. (The same effect, albeit less pronounced, is achievable through the crystal-face engineering in titania photoanode [7,8]). The second strategy is more versatile in view of a large palette of redox couples to be selected to this purpose. The most often used redox mediator is the $I^-/I_3^-$ couple, but its too negative redox potential (against that of the dye), complicated two-electron redox chemistry, corrosiveness and visible-light absorption are strong challenges for its replacement by other redox systems [1,2].

Co(III/II) complexes have quite elegantly addressed some of these problems: they are outer-sphere one-electron redox systems with more positive and easily tunable redox potentials, providing larger $V_{OC}$ [3,9-18]. Another advantage is the fast charge-transfer kinetics of Co(III/II) at graphene electrodes, i.e. platinum is no more necessary as a counterelectrode catalyst [13-18]. Indeed, the stepwise enhancement of the DSCs' efficiency to 12% [10], 13% [11] up to the current record (14.3%) [12] was achieved through the use of Co-complexes [10-12] and the graphene-based counterelectrodes [11,12]. Nevertheless, Co(III/II) complexes also bring some issues associated with their stability [19] slow mass transport in the electrolyte solution [20] and large internal reorganization energy between the high-spin $d^7$ and low-spin $d^6$ states (which costs additional driving force for dye regeneration) [21].

The limitations caused by the reorganization energy are minimized with alternative Cu-redox mediators. They are based on tetra-coordinated Cu(II/I) complexes, with distorted tetragonal symmetry, such as [Cu(dmp)$_2$]$^{2+/+}$; dmp = bis(2,9-dimethyl-1,10-phenantroline) [22–24]. By optimization of the cell components (particularly the dyes) the efficiency of Cu-mediated DSCs quickly jumped from the initial value of 1.4% [23] to 7% [22], 8.3% [24] up to the current record of 10.3% [25]. (For a review of copper complexes in DSCs see [26]). Interestingly, the 8% efficiency was still present even in some unsealed cells, from which the solvent slowly evaporated turning the liquid electrolyte solution into a solid residuum (the so called 'Zombie Cells') [27]. This residuum served as a solid hole-conductor, but the mechanism of charge transport is still unclear.

Very recently, Saygili e al. [25] introduced two new tetra-coordinated Cu(II/I)-complexes with bipyridine ligands, viz. 6,6'-dimethyl-2,2'-bipyridine (dmby) and 4,4',6,6'-tetramethyl-2,2'-bipyridine (tmby) with TFSI (trifuloromethylsufonylimide) counter-ions. These new complexes not only offered comparable efficiencies to the best Cu-dmp-mediated devices (10.0-10.3%) but an interesting conclusion was associated with the Cu(dmby)$_2$$^{2+/+}$ exhibiting an extremely positive redox potential of 0.97 V vs. SHE, but still working well with the Y123 dye (redox potential of 1.07 V vs. SHE). In other words, this complex confirmed the unexpected finding, that a driving force as small as 0.1 eV is still sufficient for dye regeneration at a quasi-quantitative level in these DSCs [25].

One of the fundamental problems associated with the tetra-coordinated tetrahedral Cu-complexes is the soft nature of their coordination sphere and its significant dependence on the oxidation states of Cu. More specifically, there is a strong preference of Cu(II) but not Cu(I) for less-polarizable- and more Lewis-basic ligands. This effect was reported by Hoffeditz et al. [28] using the tetradentate ligand, 1,8-bis(2,2'-pyridyl)-3,6-dithiaoctane (PDTO). Their Cu(II)(PDTO) was very sensitive to substitution reactions with TBP which is currently used in almost every electrolyte solution for DSCs (see above) [6]. In acetonitrile (AN) solution, various 4-, 5- and 6-coordinated complexes, [Cu(TBP)$_x$(AN)$_y$]$^{2+}$ were detected and even the counterion (CF$_3$SO$_3$$^-$; triflate) was found to enter the coordination sphere in [Cu(TBP)$_4$(triflate)$_2$] complex [28]. The complicated coordination chemistry was implicitly mentioned also with the state-of-art DSCs using Cu(dmby)$_2$$^{2+}$ and Cu(tmby)$_2$$^{2+}$ mediators:

cyclic voltammograms were dependent on the TBP addition, and the Cu(II) complexes had significantly different cyclic voltammogram than the Cu(I) complexes [25].

In this work, we addressed the 'TBP-effect' in more detail, and found a strikingly strong influence of TBP on the mass-transport and charge transfer kinetic at the counterelectrode with three different catalysts: PEDOT, Pt and graphene. The latter catalyst was used for the first time with promising results. This finding contrasts favorably with the earlier report that carbon black was not good for the $Cu(dmp)_2^{2+/+}$ couple unless the carbon was platinized [22]. Furthermore, we solved, in this work, the stability issues of Cu(II)-bipyridine species by development of a novel electrochemical synthetic protocol to get the electrochemically clean $Cu(dmby)_2^{2+}$ and $Cu(tmby)_2^{2+}$ complexes in stoichiometric yield.

## 2. Experimental Section
Chemicals

The $Cu(dmp)_2TFSI_{2/1}$, $Cu(dmby)_2TFSI_{2/1}$ and $Cu(tmby)_2TFSI_{2/1}$ complexes were available from our earlier work [25]. Briefly, all the Cu(I) complexes were synthesized by the reaction of CuI with the corresponding ligand. The $Cu(dmp)_2TFSI_2$ was produced by oxidation $Cu(dmp)_2TFSI$ with $NOBF_4$. The $Cu(dmby)_2TFSI_2$ and $Cu(tmby)_2TFSI_2$ were prepared directly by a reaction of $CuCl_2$ with the corresponding ligand [25]. $Co(bpy)_3(TFSI)_{3/2}$ were from Dyenamo, AB. Other chemical were from Aldrich or Merck, and used as received from the supplier. LiTFSI and $NOBF_4$ were handled in a glove box under Ar.

Electrode materials

FTO glass (TEC 15 from Libbey-Owens-Ford, 15 Ohm/sq) was ultrasonically cleaned in a detergent solution (Deconex) followed by sonication in water, ethanol and acetone (30 min each). The PEDOT-coated FTO electrodes were fabricated by an electrochemical deposition from EDOT as detailed in [25]. Platinized FTO was prepared by deposition of 5 µL/cm$^2$ of 10 mM $H_2PtCl_6$ in 2-propanol and calcination at 400$^o$C for 15 minutes. Stacked graphene platelet nanofiber acid washed (SGNF) was from ABCR/Strem (identical catalyst was used in some earlier works on Co-mediators [11,18,20]). SGNF was dispersed in water by sonication and the dispersion was left overnight to separate larger particles by sedimentation.

The supernatant containing about 1 mg/mL was stable for several days without marked sedimentation. Single-layer graphene oxide (GO) was from Cheap Tubes, Inc. It was dispersed in water by sonication to a concentration of 1 mg/mL. The precursor for composite electrodes was prepared by mixing the SGNF dispersion with GO solution to a desired proportion of both components. The optimum proportion was 80 wt% of SGNF and 20 wt% of GO; thus made electrode is abbreviated GONF80 [18]. The graphene films on FTO were deposited from aqueous solutions by air-brush spraying over a warm substrate (ca. 100$^o$C). The amount of deposited carbon was adjusted by the time of spraying, and was quantified by measurement of optical density (See Figure S1, Supporting Info). Consistent with our previous works [15,16,18,29], the optical transmittance at a wavelength of 550 nm, $T_{550}$ served as a parameter characterizing the films. The final graphene-based films on FTO had the $T_{550}$ values between 75 and 83 %; PEDOT had ca. 60 % and Pt around 95 % transmittance.

The symmetrical dummy cell was fabricated from two identical FTO-supported electrodes which were separated by Surlyn (DuPont) tape as a seal and spacer. The sheet edges of FTO were coated by ultrasonic soldering (Cerasolzer alloy 246, MBR Electronics GmbH) to improve electrical contacts. The distance between electrodes was measured by a digital micrometer. The cell was filled with an electrolyte through two holes in one FTO support which was finally closed either by Kapton foil or by a Surlyn seal. Electrolyte solution was identical to that used in solar tests reported earlier (L is the corresponding ligand) [25]: 0.2 M CuL$_2$TFSI + 0.04 M CuL$_2$(TFSI)$_2$ + 0.1 M LiTFSI + 0.5 M TBP (the last component was avoided in certain tests, see the main text) in propionitrile or acetonitrile. Electrolyte solution for the reference Co-mediator was: 0.3 M Co(bpy)$_3$(TFSI)$_2$ + 0.1 M Co(bpy)$_3$(TFSI)$_3$ + 0.1 M LiTFSI + 0.5 M TBP in propionitrile.

Methods

Electrochemical measurements were carried out using Autolab PGstat-30 equipped with the FRA module (Ecochemie). A standard Ar-purged one-compartment three-electrode cell was used for cyclic voltammetry in solution. Preparative electrolysis was carried out potentiostatically at platinized FTO electrode using a divided cell, in which the counterelectrode was separated by two frits from the working-electrode. Electrochemical impedance data were processed using Zplot/Zview software. The impedance spectra were acquired in the frequency range from 100 kHz to 0.1 Hz, at 0 V bias voltage, the modulation amplitude was 10 mV. Ag/AgCl/saturated LiCl (ethanol) served as reference electrode and glassy carbon or platinum were the working electrodes. The optical spectra were measured by the Perkin Elmer Lambda 1050 spectrometer with integrating sphere in transmission mode. The reference spectrum was air. Experiments on dye-sensitized solar cells were carried out as in our earlier work [25]. Briefly, FTO-supported, $TiO_2$ double-layer photoanode (5 μm nanocrystalline + 5 μm scattering layers) was prepared from Dyesol precursors and sensitized with Y123 dye (Dyenamo AB; for chemical formula see Fig. S2 in Supporting Info). The DSCs were assembled with 25 μm Surlyn (Dupont) spacer and sealant. The electrolyte solution was identical to that used for dummy cells (see above). The DSC's current-voltage characteristics were obtained by using a 450 W xenon light source (Osram XBO 450, Germany) with a filter (Schott 113). The light power was regulated to the AM 1.5G solar standard by using a reference Si photodiode equipped with a color-matched filter (KG-3, Schott) to reduce the mismatch between the simulated light and AM 1.5G to less than 4% in the wavelength region of 350–750 nm. The differing intensities were regulated with neutral wire mesh attenuator. The applied potential and cell current were measured using a Keithley model 2400 digital source meter.

## 3. Results and Discussion
### 3.1. Counterelectrode catalysts, mass transport and effect of TBP

The electrocatalytic activity of DSCs' cathodes and mass transport in the electrolyte solution are conveniently studied using symmetrical dummy cells [13]. The most popular cathode catalyst, which was employed in all champion devices as well as in the 'Zombie Cells' is PEDOT [24,25,27], sometimes also platinum was used [22,23,28]. Figure 1 (top charts) presents cyclic voltammograms and electrochemical impedance spectra (the latter were fitted to the equivalent circuit, model 1, Figure S3, Supporting Info) of symmetrical dummy cells. In these cells, the electrolyte solution with $Cu(dmp)_2^{2+/+}$ is sandwiched between two identical FTO-supported PEDOT electrodes. For comparison, the bottom charts display the corresponding data for the $Co(bpy)_3^{3+/2+}$ couple in analogous dummy cells.

The most striking difference between Cu- and Co-based mediators is the influence of TBP addition. While the electrochemical behavior of $Co(bpy)_3^{3+/2+}$ is invariant with the TBP addition, the Cu(dmp)$_2^{2+/+}$ is not. This is further illustrated by standard cyclic voltammograms, measured on a Pt-electrode in the conventional three-electrode set-up. Figure S4 (Supporting Info) shows that the voltammograms of Co(III) and Co(II) occur at the same formal potential, independent of TBP, solely the diffusion currents are shifted in accordance with the presence of the particular diffusion-limiting species. In contrast, the same experiment with Cu(dmp)$_2^{2+/+}$ reported in [25] indicated still the same formal potentials for the Cu(I)-dmp and Cu(II)-dmp counterparts (with expectedly shifted diffusion currents), but there was considerable downshift of the formal potentials for both Cu(I) and Cu(II) species caused by the addition of TBP.

The effect of TBP addition to Cu(dmp)$_2^{2+/+}$ can be more clearly investigated on symmetrical dummy cells (Figure 1, top charts). Cyclic voltammogram in the TBP-free electrolyte solution confirms that the charge transfer and diffusion are quite fast. The inverse slope of a voltammogram at the potential of 0 V characterizes the catalytic activity of an electrode; it is actually the overall cell resistance ($R_{CV}$) that can be attained at low current densities [15]. The found $R_{CV}$ (Fig. 1) is ca. 9 Ω·cm$^2$ both for the Cu- and Co-based mediators. Also the limiting diffusion currents, $j_L$ are comparable (≈20 mA/cm$^2$) though they are also dependent on the dummy cell spacing (δ):

$j_L = 2nFcD/\delta$ (1)

($n$ = 1 is the number of electrons, $F$ is the Faraday constant, $c$ is the concentration of diffusion-limited species (Cu(dmp)$_2^{2+}$ or Co(bpy)$_3^{3+}$) and $D$ is the diffusion coefficient. Assuming the spacing δ ≈ 30 μm (by the Surlyn foil) then the $D$ equals ca. 6·10$^{-6}$ cm$^2$/s or 3·10$^{-6}$ cm$^2$/s for the Cu(dmp)$_2^{2+}$ or Co(bpy)$_3^{3+}$, respectively in the TBP-free propionitrile solutions. While the found diffusion coefficient for the Co-mediator is comparable to the literature values for similar systems [9,15,18,24,30], the coefficient of Cu(dmp)$_2^{2+}$ is somewhat smaller than reported from rotating-disc experiments (25·10$^{-6}$ cm$^2$/s [24] or 14.4·10$^{-6}$ cm$^2$/s [25]). Nevertheless, the cited works [24,25] found that the diffusion coefficient of Cu(dmp)$_2^{2+}$ was larger than that of Co(bpy)$_3^{3+}$ by a factor of ca. 2, which is consistent with our data, see above.

The faster diffusion of $Cu(dmp)_2^{2+}$ was ascribed to a smaller size of this ion [24] but in view of the instability of the coordination sphere [28], this argument should be used with care. We can also speculate that very fast electron self-exchange rate constant ($k_{ex} = 23$ M$^{-1}$s$^{-1}$, which is 1000 times larger than that of the homologue phenanthroline complex, $Cu(phen)_2^{2+/+}$) [23] could contribute to charge transfer by electron-hopping between redox molecules. For the center-to center distance $\delta_{cc}$, the charge percolation occurs by hopping without physical motion of molecules in the electrolyte solution (Dahms-Ruff mechanism [31,32]). According to this model, the observed diffusion coefficient actually corresponds to the sum of the mass transport by physical displacement of molecules, $D_{mass}$ and the electron-hopping contribution:

$$D = D_{mass} + \frac{k_{ex} c \delta_{cc}^2}{6} \qquad (2)$$

We can estimate $\delta_{cc} \approx 2$ nm from the used concentrations (see Experimental Section), from which the hopping contribution to the diffusion coefficient is of the order of $10^{-13}$ cm$^2$/s, i.e. negligible for pure $Cu(dmp)_2^{2+/+}$ complexes at the concentrations employed in the redox electrolytes. However, if we consider the ligand exchange and/or the formation of larger aggregates, then the $k_{ex}$ can be significantly higher. For instance, the electron-transfer in $[Cu(dmp)_2(AN)]^{2+}/[Cu(dmp)_2]^+$ has $k_{ex} = 10^3$ M$^{-1}$s$^{-1}$ and values of ca. $10^5$-$10^6$ M$^{-1}$s$^{-1}$ were found for Cu complexes with certain proteins [33].

More detailed information follows from the evaluation of electrochemical impedance spectra (EIS). The spectra of dummy cells with simple catalytic electrodes (PEDOT, Pt of SGNF) were fitted to the model 1 equivalent circuit (Figure S3, Supporting Info), but the spectra of composite graphene electrodes (GONF80) need to be fitted to a more complicated circuit, model 2 [14,18]. In all cases, the catalytic activity is evaluated from the high-frequency part of the spectrum by a charge transfer resistance, $R_{CT}$ which is in series to the constant phase elements (CPE, CPE$_1$, CPE$_2$). The use of CPE is necessary, due to the deviation from the ideal capacitance by the electrode inhomogeneity (roughness) [13]. The impedance of constant phase element equals:

$$Z_{CPE} = B(i\omega)^{-\beta} \qquad (3)$$

where ω is the frequency and $B$, $β$ are the frequency-independent parameters of the CPE ($0 ≤ β ≤ 1$; the corresponding parameters are 'CPE-T'= $B$ and 'CPE-P = $β$).

The fitted $R_{CT}$ values scale inversely with the exchange current density, $j_0$ at the cathode [13]:

$$j_0 = \frac{RT}{nFR_{CT}} = Fk_0(c_{ox}^{1-\alpha} \cdot c_{red}^{\alpha}) \tag{4}$$

$R$ is the gas constant, $T$ is temperature, $n=1$ is the number of electrons, $k_0$ is the formal (conditional) rate constant of the electrode reaction, $c_{ox}$ and $c_{red}$ are the concentrations of oxidized and reduced mediator, respectively and α is the charge-transfer coefficient ($α ≈ 0.5$). In good DSCs, the value of $j_0$ should comparable to the short-circuit photocurrent density at 1 sun illumination.

The low-frequency part of the spectrum is dominated by the Warburg impedance. It is modelled by a finite-length element $W_s$ with the parameters 'W$_s$-R'=$R_W$, 'W$_s$-T'=$T_W$ and 'W$_s$-P' = 0.5 [4,34]. The finite length Warburg diffusion impedance is expressed as:

$$Z_W = \frac{R_W}{\sqrt{iT_W\omega}} \tanh\sqrt{iT_W\omega} \quad ; \quad T_W = \frac{\delta^2}{D} \tag{5}$$

The fitted values are collected in Table 1.

The fast charge-transfer on PEDOT electrode is confirmed by a very small value of $R_{CT}$ of 0.07 Ω·cm$^2$ which is even better than that of the control Co-system (0.20 Ω·cm$^2$). Other tested cathode catalysts (Pt and SGNF) provide similarly good $R_{CT}$ values (0.37 and 0.27 Ω·cm$^2$) which leads to a somewhat unexpected finding (cf. ref. [22]) that graphene cathode is electrocatalytically highly active for Cu(dmp)$_2^{2+/+}$, outperforming even the activity of platinum (Figure 2 and Table 1). Also the Warburg diffusion resistance, $R_W$ is small: 4.6-5.6 Ω·cm$^2$ for all three catalysts tested. It is comparable to 4.0 Ω·cm$^2$ found for Co-

bipy@PEDOT, the latter being reasonably similar to the literature value (2.9 Ω·cm$^2$) for an analogous Co-mediated system [30].

Unfortunately, these promising properties of Cu(dmp)$_2$$^{2+/+}$ are strongly impaired by TBP, whereas with the Co(bpy)$_3$$^{3+/2+}$ +TBP solutions, the unperturbed charge transfer and diffusion are still observed in the control devices. For all the three catalysts tested, the $R_{CT}$ and $R_W$ values in the TBP-containing solutions of Cu(dmp)$_2$$^{2+/+}$ increase ca. 10 and 5 times, respectively (Table 1). The fact that both the catalytic activity and diffusion rate are deteriorated by TBP is reminiscent of the findings by Kim et al. [4] for the traditional system, i.e. I$^-$/I$_3$$^-$ solution in the Pt@FTO/Pt@FTO dummy cells. In this case, the enhancement of TBP concentration from 0.5 M to 2.5 M caused $R_{CT}$ and $R_W$ to increase by a factor of 3 and 2, respectively, which was attributed to the adsorption of TBP on Pt (impairing access of redox molecules to the catalytic surface) and enhanced viscosity of the solution (impairing diffusion) [4]. In the case of the Cu(dmp)$_2$$^{2+/+}$ mediator, we need to consider also qualitatively different effects than adsorption and viscosity (though viscosity is also at play, as seen from the $R_W$ values for our AN and PN solutions in Table 1).

This hypothesis is supported by the fact that (a) the Cu(II) species is the low-concentration-, i.e. the diffusion-controlling ion in the system and (b) the same Cu(II) species is prone to changes of its coordination sphere. These changes include three different effects: (b$_1$) variations of coordination geometry, (b$_2$) increase of coordination number (from 4 to 6) and (b$_3$) ligand substitution by TBP, solvent and/or counterions [28]. The Cu(II) species in solution likely becomes more bulky and hydrophobic due to these changes, which accounts for the sluggish diffusion and charge-transfer. Nevertheless, the final $R_{CT}$ and $R_w$ values in TBP-containing solutions are still not critical for the use in solar cells, i.e. the DSCs using these solutions still exhibit the short circuit photocurrent of ca. 14 mA/cm$^2$ and efficiency of ca. 10% at 1 sun illumination [25]. From Eq. (4) the current of 14 mA/cm$^2$ translates into the $R_{CT}$ value of 1.8 Ω·cm$^2$ if we adopt the condition that the exchange current density at the counterelectrode ($j_o$) equals the short-circuit photocurrent. Hence, the good counterelectrode should exhibit $R_{CT}$ < 1.8 Ω·cm$^2$. The benchmark is met in some of the actual counterelectrode materials and electrolyte solutions used in solar cells (Table 1). Nevertheless, this analysis highlights a strong *need for replacing TBP in Cu-mediated DSCs by another additive which does not decrease the charge-transfer and mass transport rates.* There are some other $V_{OC}$- improving analogues of TBP, e.g. N-butylbenzimidazole (NBB) or other N-containing heterocyclic compounds [6]. Unfortunately, our preliminary experiments with NBB were not successful: this additive has similarly perturbing effect like TBP (Fig. S5, Supporting Info). Hence, the replacement of TBP (NBB) by another ancillary species is still a challenge.

The effects, which were observed for $Cu(dmp)_2^{2+/+}$ are qualitatively reproduced also for the Cu-bipyridine mediators. Figure 3 shows example data for $Cu(tmby)_2^{2+/+}$ solutions at two different graphene electrodes. For practical reasons, the composite graphene electrode (GONF80) is preferable to pure SGNF, because the former provides considerably better adhesion of the catalyst to the substrate (FTO) [18]. On the other hand, pure SGNF [11,20] is known to provide higher electrocatalytic activity than the composites with graphene oxide [18] for Co-mediated systems, and this finding is reproduced for our Cu-mediators, too (Table 1). However, in this particular case (GONF80 catalyst), the impedance spectra must be fitted to a more complicated circuit with two time constants, model 2 (Fig. S3, Supporting Info) [14,18]. In most cases (PEDOT, SGNF, GONF80 catalysts and AN, PN solutions) we observe reasonable catalytic activity with $R_{CT}$ around 3 $\Omega \cdot cm^2$, but interestingly, platinum electrode is considerably less active both in the PN and AN solutions of $Cu(tmby)_2^{2+/+}$. We have so far no explanation for the difference; but qualitatively note that Pt is the least active catalyst also for $Cu(dmp)_2^{2+/+}$.

## 3.2. The 'Zombie Effect'

One of the interesting challenges in the Cu-mediated DSCs is the formation of the so called "Zombie Cells' [27]. They denote solar cells, in which the solvent slowly evaporated from the electrolyte solution through imperfect sealing, but they still worked surprisingly well. This was intuitively ascribed to the fact that the solid residuum after evaporation of the solvent behaves like a hole-conductor. The corresponding solid-state DSC exhibited the largest solar conversion efficiency (8.2 % vs. 5.6% for a control device with *spiro*-OMeTAD) [27]. The 'Zombie Effect' was so far observed only for full DSCs with titania photoanode. Here we reproduced several zombies also in the symmetrical PEDOT/PEDOT dummy cells.

Figure 4 shows some typical data. In general, the experiments were poorly reproducible, but in most cases we observed that our zombies exhibited improved charge transfer kinetics, but a larger diffusion resistance compared to that of the cells with the parent liquid electrolyte solution. The evaporation at room temperature and pressure leads to intermediate stages in which the solid hole-conductor and liquid electrolyte seem to work in parallel. Slow evacuation provides presumably a solid hole-conductor in which the progressive improvement of charge transfer kinetics goes along with a slowdown of the charge transport. Presumably, the hopping charge-transfer (Dahms-Ruff mechanism, Eq. 2) which is less important in the liquid solutions of pure complexes, becomes the main hole-transport process in the solid conductor of the zombies.

## 3.2. Preparation of electrochemically clean Cu(II)-bipyridine complexes

Our previous work [25] concluded that the $Cu(tmby)_2^{2+}$ and $Cu(dmby)_2^{2+}$ species did not behave electrochemically like the corresponding Cu(I) counterparts, but the $Cu(dmp)_2^{2+}$ complex did. More specifically, the Cu(II)-bipyridine complexes exhibited strange two-waves voltammograms with downshifted formal potentials, clearly pointing at more complicated coordination chemistry in these complexes, but this problem remained open in the cited work [25].

To address this peculiarity, we first recall the fact the 'neat' Cu(II)-dmp species is synthesized from the parent Cu(I)-dmp by chemical oxidation with $NOBF_4$, while the 'non-neat' Cu(II)-bipyridine is made by direct synthesis from $CuCl_2$ and the corresponding ligand. This outlines a logical strategy to test the chemical oxidation of Cu(I)-bipyridine as well. Figure 5 shows that, indeed, the chemical oxidation of $Cu(tmby)_2^+$ successfully provided the electrochemically clean Cu(II)-bipyridine. The same result was obtained for $Cu(dmby)_2^+$ (data not shown). The oxidized species exhibited the expected voltammogram with identical formal potentials and only with cathodically shifted plateau of diffusion currents. Hence, the reactions which are schematically depicted as follows:

$Cu(tmby)_2^+ + NOBF_4 \rightarrow Cu(tmby)_2^{2+} + BF_4^- + NO$ \hfill (6a)

$Cu(dmby)_2^+ + NOBF_4 \rightarrow Cu(dmby)_2^{2+} + BF_4^- + NO$ \hfill (6b)

are readily applicable for the bipyridine complexes like for the phenanthroline complexes. However, closer inspection of Fig. 5 reveals that the reactions (6a, 6b) are by far not stoichiometric, i.e. more than 1 equivalent of the oxidant is needed to complete the oxidation. Furthermore, the reaction proceeds quite slowly (progressing during 10 hours of aging the reaction mixture) and is accompanied by partial oxidative destruction of the starting substance which manifests itself by decreasing voltammetric currents at the final stages of the reaction.

All these problems are elegantly avoided if the oxidation is carried out electrochemically instead of chemically. A preparative electrolysis of Cu(I)-bipyridine in divided cell and potentiostatic mode leads to clean Cu(II)-bipyridine products in stoichiometric yield and free from any oxidative destruction. Figure 6 shows the relevant data for $Cu(tmby)_2^+$ oxidation. Obviously, each portion of the passed charge (E1-E5) caused the relevant oxidative transformation. Identical result was obtained also for $Cu(dmby)_2^+$ (Fig. S6 in Supporting Info). The speed of preparative electrolysis depends on the stirring rate, electrode area and the applied potential, but the reaction is generally faster than chemical oxidation and does not require any harmful reactants ($NOBF_4$).

To illustrate the differences between our product of preparative electrolysis and the product of standard chemical synthesis of $Cu(tmby)_2^{2+}$ or $Cu(dmby)_2^{2+}$ (which were used in [25]) we compare the relevant voltammograms in the right chart of Fig. 6 and in Fig. S6. Dashed black line shows the cyclic voltammogram of the latter species highlighting its complicated electrochemistry. This is obviously due to changes in the coordination sphere of Cu(II)-bipyridine complexes during their direct synthesis from $Cu^{2+}$ and the corresponding ligand. On the other hand, the preparative electrolysis through anodic oxidation of Cu(I)-bipyridine species is an ideal protocol towards electrochemically clean Cu(II)-bipyridine counterparts of the redox couple.

Finally, we tested our newly prepared electrochemically-clean redox mediators in dye-sensitized solar cells. Figure S7 (Supporting Info) presents example IV-curves for the $Cu(dmby)_2^{2+/+}$-mediator in acetonitrile solution, and Table S1 (Supporting Info) collects the corresponding DSCs' parameters. This study upgrades our earlier work [25] in which the impure $Cu(dmby)_2^{2+}$ was used (from direct chemical synthesis, 'as-rec'). The performance of electrochemically-made $Cu(dmby)_2^{2+}$ is promising: the corresponding $Cu(dmby)_2^{2+/+}$-mediated DSC shows better efficiency as referenced to a control device with the as-rec

Cu(tmby)$_2^{2+/+}$-mediator (Fig. S7 and Table S1). We may note that opposite performance metrics was reported for Cu(dmby)$_2^{2+}$ vs. Cu(tmby)$_2^{2+}$ which were grown by classical chemical synthesis [25].

**Conclusions**

Three different Cu-based redox mediators for high-voltage Cu(dmp)$_2^{2+/+}$, Cu(dmby)$_2^{2+/+}$ and Cu(tmby)$_2^{2+/+}$ were studied on various electrodes and in variety of electrolyte solutions using cyclic voltammetry and impedance spectroscopy on symmetrical dummy cells. Graphene-based catalysts provide comparably high activity to PEDOT, which was so far a standard counterelectrode catalyst in Cu-mediated DSCs. Both graphene and PEDOT outperform the electrocatalytic activity of platinum.

The charge-transfer kinetics and the diffusion rate of Cu-mediators depend dramatically on the presence of TBP, which is a standard additive to electrolyte solutions for dye-sensitized solar cells. While TBP has almost no effect on Co(bpy)$_3^{3+/2+}$ mediators, significant slowdown of electrode kinetics and diffusion rate is observed in the case of Cu-mediators. This finding is ascribed to a sensitivity of the coordination sphere of Cu(II)-species to structural and substitutional changes. Though these changes do not impair critically on the function of solar cells, the replacement of TBP is an upfront challenge for future studies.

The 'Zombie Cells' made from symmetrical PEDOT/PEDOT devices exhibit enhanced charge transfer rate end enhanced diffusion resistance during progressive evaporation of the solvent from the electrolyte solution. The charge transport mechanism was discussed in terms of high electron self-exchange, possible percolation-hopping in the used Cu-complexes, and the ligand-change reactions in the Cu(II)-counterparts of the redox couple.

The instability of coordination sphere of the Cu(II)-species manifest itself also by the complications during their synthesis from Cu$^{2+}$ and the corresponding ligand. These problems are avoided if the synthesis is carried out by chemical (by NOBF$_4$) or electrochemical oxidation of the parent Cu(I) species, which is stable. While the chemical oxidation is poorly controlled and complicated by a partial destruction of the product, the electrochemical route provides electrochemically clean Cu(II)-species rapidly, in stoichiometric yield and without using harmful chemical oxidants. The superior performance of electrolyte solution based on electrochemically-clean Cu(II)-species manifests itself in dye-sensitized solar cells, too.

**Acknowledgment**

This work was supported by the Grant Agency of the Czech Republic (contract No. 13-07724S), by the European Union H2020 Program (no. 696656–Graphene Flagship Core1) and by the Swiss National Science Foundation project "Fundamental studies of mesoscopic devices for solar energy conversion" with project number 200021_157135/1.

**Appendix A. Supplementary data**

Supplementary data associated with this article can be found, in the online version, at ….

**Table 1**. Electrochemical parameters of the studied cathode materials in symmetrical dummy cells obtained from fitting of impedance spectra

| Electrode | Electrolyte solution | $R_s$ ($\Omega.cm^2$) | $R_{CT}$ ($\Omega.cm^2$) | $R_W$ ($\Omega.cm^2.s^{-1/2}$) | CPE:B ($S.s^\beta$) | CPE: $\beta$ |
|---|---|---|---|---|---|---|
| PEDOT | Co(bpy)$_3$/PN | 1.8 | 0.19 | 3.9 | $1.0 \cdot 10^{-3}$ | 0.83 |
| PEDOT | Co(bpy)$_3$/PN (TBP free) | 1.8 | 0.20 | 4.0 | $1.1 \cdot 10^{-3}$ | 0.83 |
| PEDOT | Cu(dmp)$_2$/AN | 2.4 | 2.1 | 7.1 | $4.1 \cdot 10^{-4}$ | 0.86 |
| Pt | Cu(dmp)$_2$/AN | 2.6 | 2.6 | 9.4 | $9.8 \cdot 10^{-6}$ | 0.89 |
| SGNF | Cu(dmp)$_2$/AN | 2.0 | 2.0 | 10.4 | $2.2 \cdot 10^{-5}$ | 0.91 |
| PEDOT | Cu(dmp)$_2$/PN | 1.7 | 1.8 | 22.2 | $6.1 \cdot 10^{-4}$ | 0.97 |
| Pt | Cu(dmp)$_2$/PN | 2.7 | 2.9 | 24.7 | $2.3 \cdot 10^{-5}$ | 0.89 |
| SGNF | Cu(dmp)$_2$/PN | 1.9 | 2.8 | 23.3 | $4.6 \cdot 10^{-5}$ | 0.88 |
| PEDOT | Cu(dmby)$_2$/AN | 2.5 | 3.3 | 18.9 | $5.8 \cdot 10^{-4}$ | 0.94 |
| Pt | Cu(dmby)$_2$/AN | 1.9 | 6.9 | 20.1 | $4.3 \cdot 10^{-5}$ | 0.87 |
| SGNF | Cu(dmby)$_2$/AN | 2.3 | 3.0 | 21.4 | $5.1 \cdot 10^{-5}$ | 0.87 |
| PEDOT | Cu(tmby)$_2$/AN | 2.4 | 3.0 | 22.7 | $3.6 \cdot 10^{-4}$ | 0.94 |
| Pt | Cu(tmby)$_2$/AN | 2.6 | 7.8 | 24.9 | $1.0 \cdot 10^{-5}$ | 0.89 |
| SGNF | Cu(tmby)$_2$/AN | 2.9 | 2.9 | 21.7 | $3.1 \cdot 10^{-5}$ | 0.94 |
| PEDOT | Cu(tmby)$_2$/PN | 1.8 | 3.15 | 24.7 | $3.3 \cdot 10^{-6}$ | 0.94 |
| Pt | Cu(tmby)$_2$/PN | 2.0 | 10.2 | 27.8 | $9.4 \cdot 10^{-6}$ | 0.84 |
| SGNF | Cu(tmby)$_2$/PN | 1.5 | 3.45 | 25.9 | $1.4 \cdot 10^{-5}$ | 0.85 |
| PEDOT | Cu(dmp)$_2$/PN (TBP free) | 1.8 | 0.07 | 4.6 | $3.7 \cdot 10^{-3}$ | 0.77 |
| Pt | Cu(dmp)$_2$/PN (TBP free) | 2.6 | 0.37 | 5.6 | $3.8 \cdot 10^{-6}$ | 0.72 |
| SGNF | Cu(dmp)$_2$/PN (TBP free) | 1.4 | 0.27 | 4.9 | $9.8 \cdot 10^{-4}$ | 0.55 |
| GONF80[a] | Cu(tmby)$_2$/PN | 1.2 | 1.2+1.05 | 20.5 | $(3.7+0.6) \cdot 10^{-4}$ | 0.80+0.68 |

Note: a) Fitting to equivalent circuit for composite electrodes (model 2)

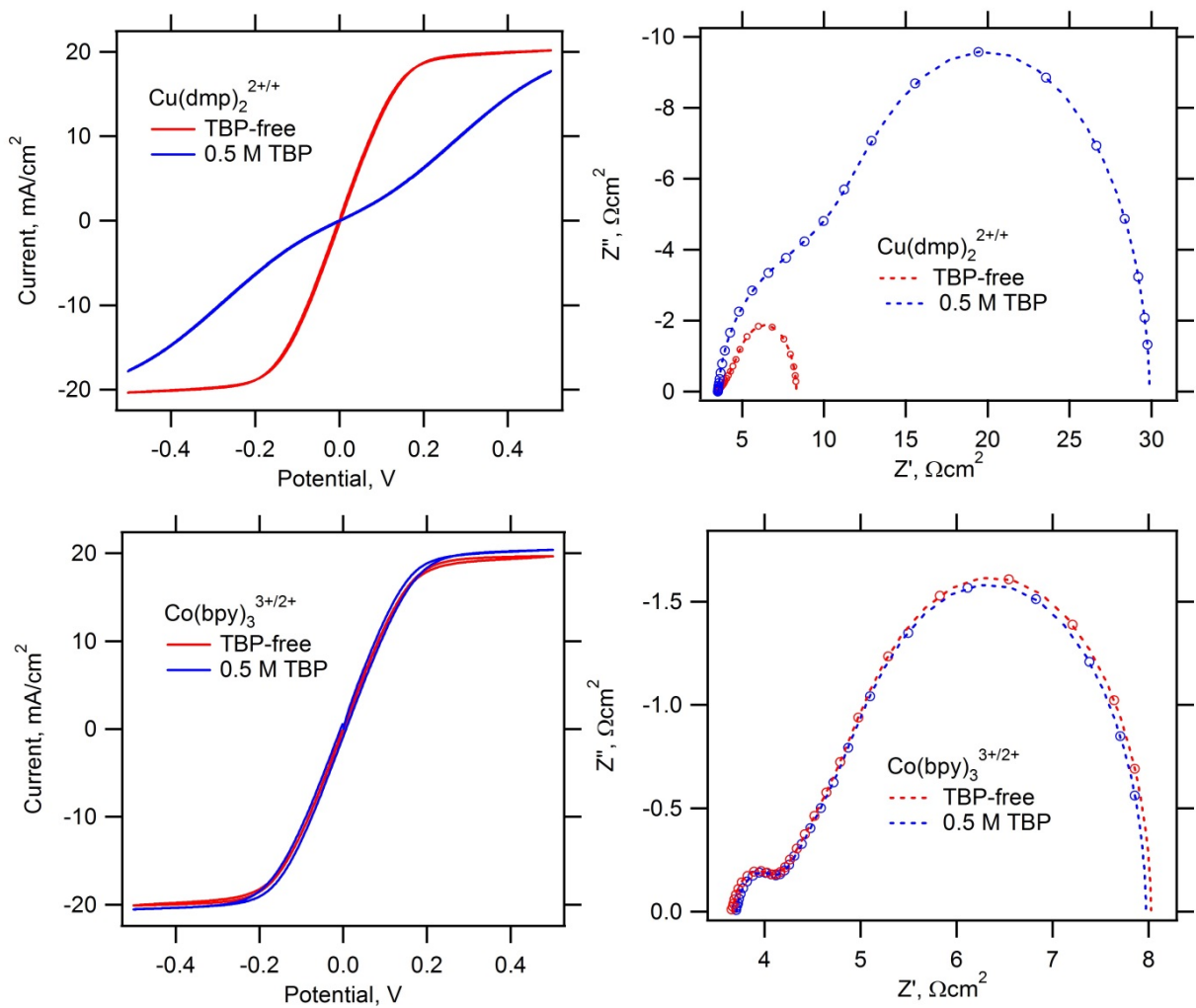

Fig. 1: Electrochemical activity of FTO supported thin film of PEDOT tested on symmetrical dummy cells. Top charts: Electrolyte solution: Co(bpy)$_3^{3+/2+}$ TFSI$_{3/2}$ in propionitrile with (blue curves) or without (red curves) the addition of 4-tert-butylpyridine. Bottom charts: Electrolyte solution Cu(dmp)$_2^{2+/+}$TFSI$_{2/1}$ in propionitrile with (blue curves) or without (red curves) the addition of 4-tert-butylpyridine. Left charts: Cyclic voltammograms, scan rate 10 mV/s. Right charts: Nyquist plots of electrochemical impedance spectra measured at 0 V from 100 kHz to 0.1 Hz. (Markers are experimental points, dashed lines are simulated fits to the equivalent circuit).

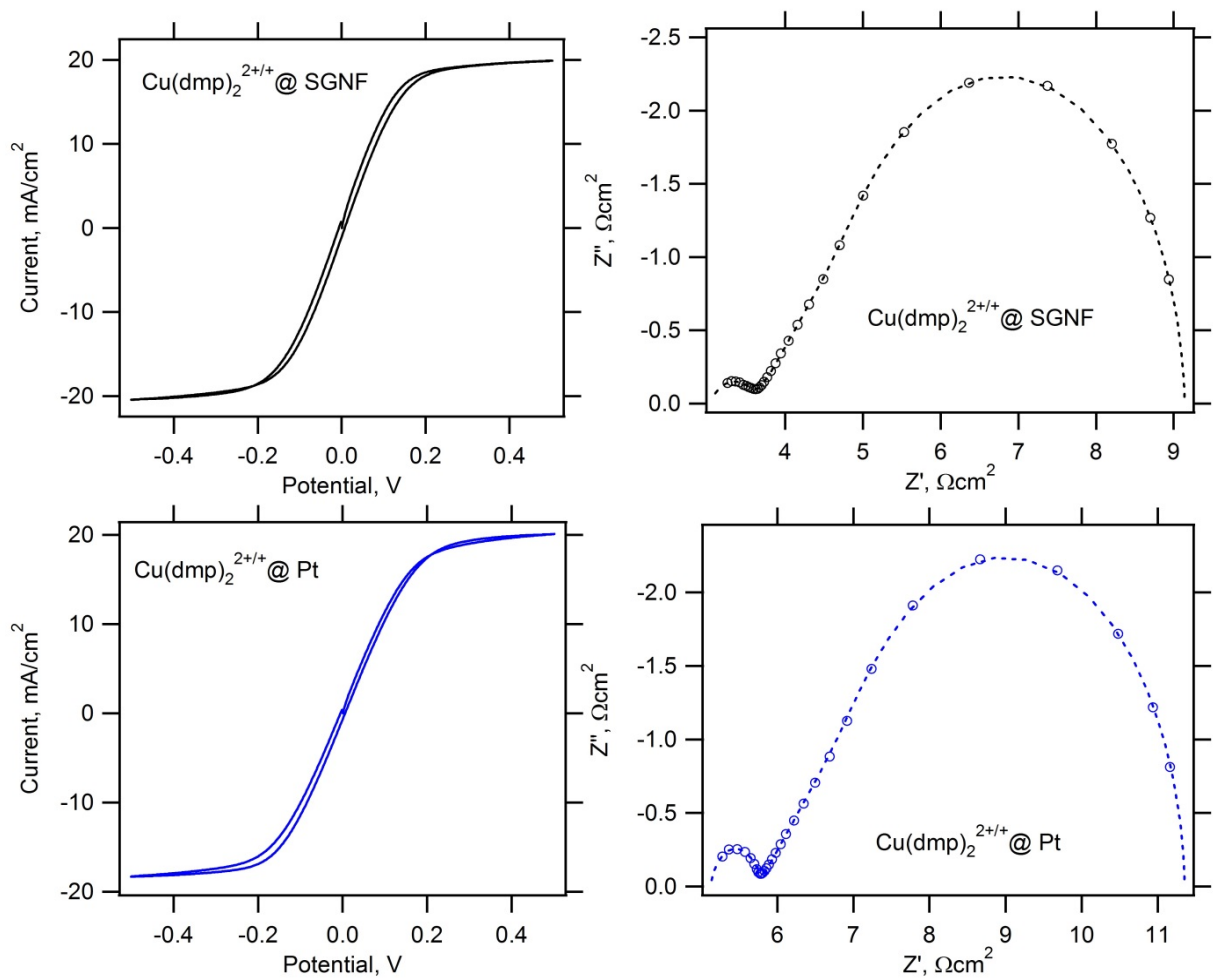

Fig. 2: Electrochemical activity of a FTO-supported thin film of stacked graphene nanofibers (SGNF, top charts) and platinized FTO (bottom charts) tested on symmetrical dummy cells. Electrolyte solution: $Cu(dmp)_2^{2+/+}$ $TFSI_{2/1}$ in propionitrile without the addition of 4-tert-butylpyridine. Left charts: Cyclic voltammograms, scan rate 10 mV/s. Right charts: Nyquist plots of electrochemical impedance spectra measured at 0 V from 100 kHz to 0.1 Hz. (Markers are experimental points, dashed lines are simulated fits to the equivalent circuit).

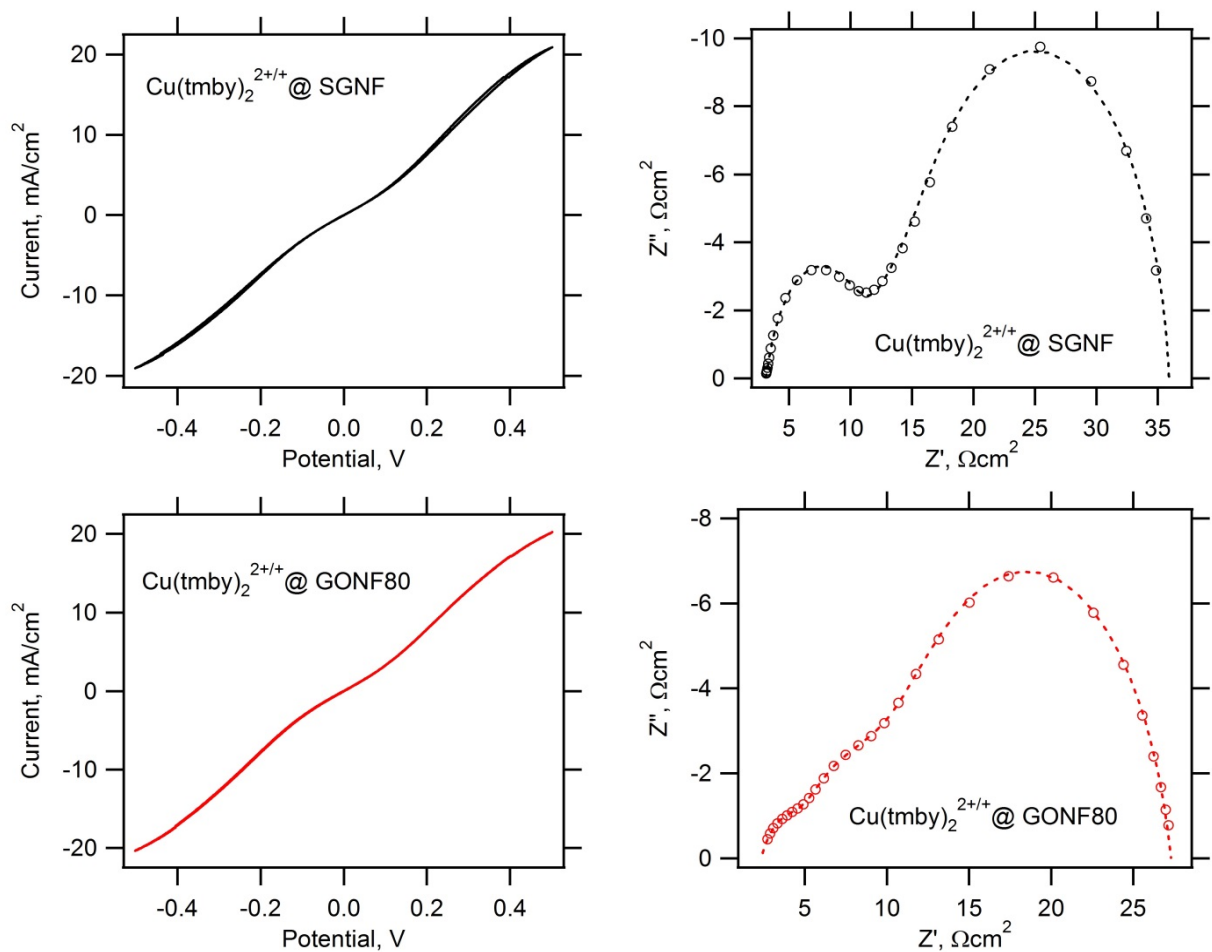

Fig. 3: Electrochemical activity of a FTO-supported thin film of stacked graphene nanofibers (SGNF; top charts) and composite of graphene oxide and SGNF (GONF80; bottom charts) tested on symmetrical dummy cells. Electrolyte solution: $Cu(tmby)_2^{2+/+}TFSI_{2/1}$ in propionitrile containing 0.5 M 4-tert-butylpyridine. Left charts: Cyclic voltammograms, scan rate 10 mV/s. Right charts: Nyquist plots of electrochemical impedance spectra measured at 0 V from 100 kHz to 0.1 Hz. (Markers are experimental points, dashed lines are simulated fits to the equivalent circuit; two different equivalent circuits were used, see text).

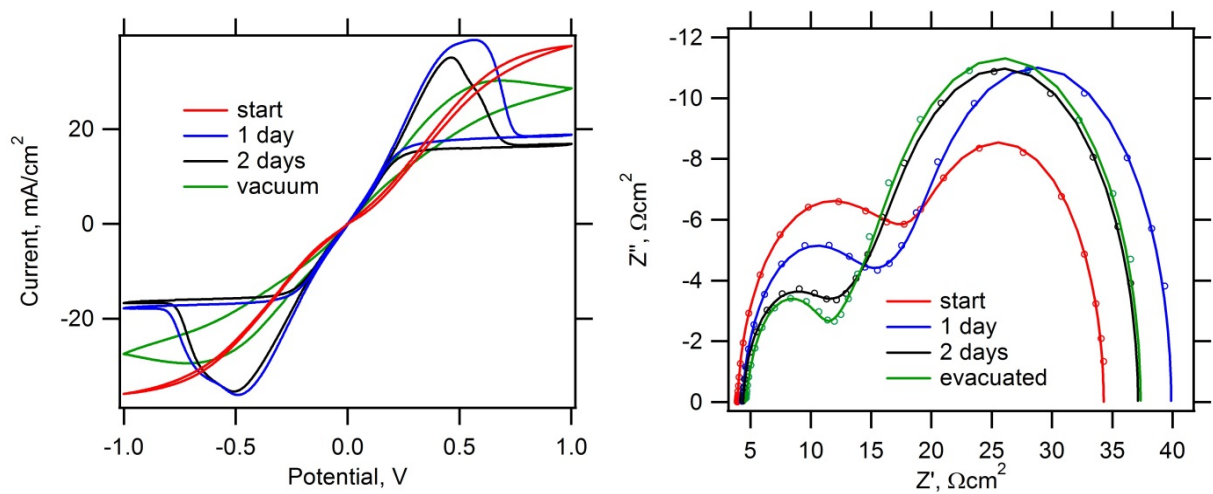

Fig. 4: Electrochemical tests of the PEDOT/PEDOT symmetrical dummy cells. Electrolyte solution: $Cu(tmby)_2^{2+/+}TFSI_{2/1}$ in propionitrile containing 0.5 M 4-tert-butylpyridine. Left charts: Cyclic voltammograms, scan rate 10 mV/s. Right charts: Nyquist plots of electrochemical impedance spectra measured at 0 V from 100 kHz to 0.1 Hz. (Markers are experimental points, lines are simulated fits). Sequential tests were made as follows: (start, red curves) – pristine closed cell with liquid electrolyte solution; (1 day, blue) – cell opened and the electrolyte solution freely evaporated for 1 day at room temperature and pressure via two filling holes; (2 days, black) – as (1) but 2 days of free evaporation; (vacuum, green) – the cell with partly evaporated solution was subjected to vacuum at room temperature.

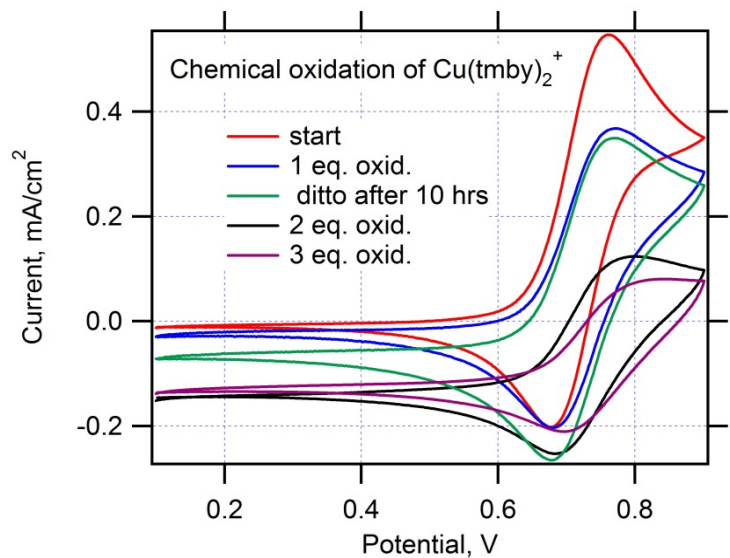

Fig. 5: Cyclic voltammograms of $Cu(tmby)_2^+TFSI$ and the products of its chemical oxidation by $NOBF_4$. Scan rate 10 mV/s. Curve 'start' (red): cyclic voltammogram of the starting 5 mM solution of $Cu(tmby)_2^+TFSI$ in 0.1 M LiTFSI + acetonitrile. The other voltammograms are for the products obtained upon addition of the given amount of $NOBF_4$ and/or equilibration by aging (see annotation in the graph).

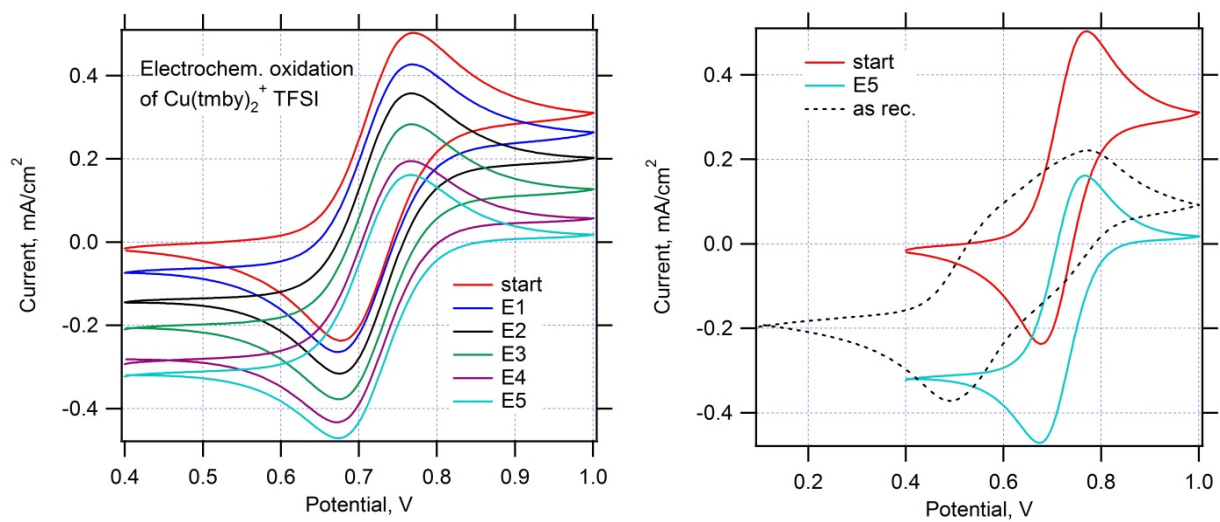

Fig. 6: Cyclic voltammograms of the products and intermediates of preparative electrolysis of $Cu(tmby)_2^+TFSI$ on Pt electrode. Scan rate 10 mV/s. Curve 'start' (red): cyclic voltammogram of the initial 5 mM solution of $Cu(tmby)_2^+TFSI$ in 0.1 M LiTFSI + acetonitrile. Voltammograms E1-E5 are for the same solution after potentiostatic oxidation at 1.1-1.4 V under stirring in divided cell. The passed charge (in equivalents of the starting amount of the Cu(I) complex) corresponds to: 0.19 eq (E1), 0.40 eq (E2), 0.60 eq (E3), 0.80 eq (E4) and 0.92 eq (E5). Right chart compares the voltammograms of the starting (start) and final (E5) complexes with the voltammogram (as rec.) of the $Cu(tmby)_2^{2+}TFSI_2$ species grown by the standard preparative protocol.